\documentclass[graybox]{svmult}

\usepackage{amsmath}
\usepackage{color}
\usepackage{cite}

\usepackage{amsfonts}
\usepackage{tikz} 
\usetikzlibrary{matrix}
\usepackage[english]{babel}
\usepackage[latin1,utf8]{inputenc}
\usepackage[T1]{fontenc}
\usepackage{ae}
\usepackage{url}

\allowdisplaybreaks[4]

\usepackage{mathptmx}       
\usepackage{helvet}         
\usepackage{courier}        
\usepackage{type1cm}        

\usepackage{makeidx}         
\usepackage{graphicx}        
\usepackage{multicol}        
\usepackage[bottom]{footmisc}

\newcommand{\bq}{\begin{eqnarray}}
\newcommand{\eq}{\end{eqnarray}}
\newcommand{\eps}{\varepsilon}

\newcommand{\slashoperator}[2]{|_{#2} #1}

\makeindex             

\begin{document}

\title*{On a class of Feynman integrals evaluating to iterated integrals of modular forms}
\author{Luise Adams and Stefan Weinzierl}
\institute{
Luise Adams \at PRISMA Cluster of Excellence, 
Institut f{\"u}r Physik, 
Johannes Gutenberg-Universit{\"a}t Mainz, 
D - 55099 Mainz, Germany, 
\email{ladams01@uni-mainz.de}, \\
Stefan Weinzierl \at PRISMA Cluster of Excellence, 
Institut f{\"u}r Physik, 
Johannes Gutenberg-Universit{\"a}t Mainz, 
D - 55099 Mainz, Germany, 
\email{weinzierl@uni-mainz.de}
}
\maketitle

\abstract{In this talk we discuss a class of Feynman integrals, which can be expressed to all orders in the
dimensional regularisation parameter as iterated integrals of modular forms.
We review the mathematical prerequisites related to elliptic curves and modular forms.
Feynman integrals, which evaluate to iterated integrals of modular forms go beyond the class of multiple polylogarithms.
Nevertheless, we may bring for all examples considered 
the associated system of differential equations by a non-algebraic transformation to an
$\eps$-form, which makes a solution in terms of iterated integrals immediate.
}

%
%

\section{Introduction}
\label{sec:intor}

\vspace{1mm}\noindent
It is an open and interesting question to which class of transcendental functions Feynman integrals evaluate.
At present, we do not have a general answer.
However, there are sub-classes of Feynman integrals for which the class of functions is known.
First of all, there is the class of Feynman integrals evaluating to multiple polylogarithms.
This covers in particular all one-loop integrals.
Starting from two-loops, there are Feynman integrals which cannot be expressed in terms of multiple polylogarithms.
The simplest example is given by the two-loop equal-mass sunrise integral\cite{Broadhurst:1993mw,Berends:1993ee,Bauberger:1994nk,Bauberger:1994by,Bauberger:1994hx,Caffo:1998du,Laporta:2004rb,Kniehl:2005bc,Groote:2005ay,Groote:2012pa,Bailey:2008ib,MullerStach:2011ru,Adams:2013nia,Bloch:2013tra,Adams:2014vja,Adams:2015gva,Adams:2015ydq,Remiddi:2013joa,Bloch:2016izu,Groote:2018rpb}.
Integrals, which do not evaluate to multiple polylogarithms are now an active field of 
research in particle physics \cite{Bloch:2014qca,Remiddi:2016gno,Adams:2016xah,Adams:2017ejb,Bogner:2017vim,Adams:2018yfj,Adams:2018bsn,Adams:2018kez,Sogaard:2014jla,Bonciani:2016qxi,vonManteuffel:2017hms,Primo:2017ipr,Ablinger:2017bjx,Bourjaily:2017bsb,Hidding:2017jkk,Passarino:2017EPJC,Remiddi:2017har,Broedel:2017kkb,Broedel:2017siw,Broedel:2018iwv,Lee:2017qql,Lee:2018ojn}
and string theory \cite{Broedel:2014vla,Broedel:2015hia,Broedel:2017jdo,DHoker:2015wxz,Hohenegger:2017kqy,Broedel:2018izr}.
In this talk we focus on a class of Feynman integrals which evaluate to iterated integrals of modular forms.
Feynman integrals of this class are associated to one elliptic curve and depend on one scale $x=p^2/m^2$.
They can be seen as generalisations of single-scale Feynman integrals evaluating to harmonic polylogarithms \cite{Vermaseren:1998uu,Remiddi:1999ew}.
We expect that all our examples are equally well expressible in terms of elliptic polylogarithms \cite{Beilinson:1994,Levin:1997,Levin:2007,Enriquez:2010,Brown:2011,Wildeshaus,Bloch:2013tra,Bloch:2014qca,Adams:2014vja,Adams:2015gva,Adams:2015ydq,Adams:2016xah,Remiddi:2017har,Broedel:2017kkb,Broedel:2017siw,Broedel:2018iwv}.
The representation in terms of iterated integrals of modular forms has certain advantages:
\begin{enumerate}
\item It combines nicely with the technique of differential equations, which by now is the main tool for solving Feynman integrals\cite{Kotikov:1990kg,Kotikov:1991pm,Remiddi:1997ny,Gehrmann:1999as,Argeri:2007up,MullerStach:2012mp,Henn:2013pwa,Henn:2014qga,Ablinger:2015tua,Adams:2017tga,Bosma:2017hrk}.
In fact, for all examples considered we are able to bring the system of differential equations into an $\eps$-form.
\item It only involves a finite number of integration kernels. The integration kernels are modular forms.
\item It allows for an efficient numerical evaluation through the $q$-expansion around the cusps \cite{Bogner:2017vim}.
\end{enumerate}
Let us also mention, that albeit an important sub-class, this class is not the end of the story.
Multi-scale integrals beyond the class of multiple polylogarithms may involve more than one elliptic curve, as seen for example
in the double box integral relevant to top-pair production with a closed top loop \cite{Adams:2018bsn,Adams:2018kez}.

\section{Periodic functions and periods}
\label{sec:periods}

\vspace{1mm}\noindent
Let us consider a non-constant meromorphic function $f$ of a complex variable $z$.
\index{period}
A period $\omega$ of the function $f$ is a constant such that
\bq
 f\left(z+\omega\right) & = & f\left(z\right)
\eq
for all $z$. 
\index{lattice}
The set of all periods of $f$ forms a lattice $\Lambda$, which is either 
\begin{enumerate}
\item trivial: $\Lambda = \{ 0 \}$,
\item a simple lattice, generated by one period $\omega$ : $\Lambda = \{ n \omega \; | \; n \in {\mathbb Z} \}$,
\item a double lattice, generated by two periods $\omega_1, \omega_2$ with $\mathrm{Im}(\omega_2/\omega_1) \neq 0$ : 
\bq
\Lambda & = & \{ n_1 \omega_1 + n_2 \omega_2 \; | \; n_1, n_2 \in {\mathbb Z} \}.
\eq
It is common practice to order these two periods such that $\mathrm{Im}(\omega_2/\omega_1) > 0$.
\end{enumerate}
An example for a singly periodic function is given by
\bq
 \exp\left(z\right).
\eq
In this case the simple lattice is generated by $\omega = 2 \pi i$.
\index{double periodic function}
An example for a doubly periodic function is given by Weierstrass's $\wp$-function.
Let $\Lambda$ be the lattice generated by $\omega_1$ and $\omega_2$
Then
\bq
 \wp\left(z\right)
 & = & 
 \frac{1}{z^2} + \sum\limits_{\omega \in \Lambda \backslash \{0\}} \left( \frac{1}{\left(z+\omega\right)^2} - \frac{1}{\omega^2} \right).
\eq
$\wp(z)$ is periodic with periods $\omega_1$ and $\omega_2$.
Of particular interest are also the corresponding inverse functions. These are in general multivalued functions.
In the case of the exponential function $x=\exp(z)$,
the inverse function is given by
\bq
 z 
 & = & 
 \ln\left(x\right).
\eq
The inverse function to Weierstrass's elliptic function $x=\wp(z)$ is an elliptic integral given by
\bq
 z 
 & = &
 \int\limits_x^\infty \frac{dt}{\sqrt{4t^3-g_2t-g_3}}
\eq
with
\bq
 g_2 = 60 \sum\limits_{\omega \in \Lambda \backslash \{0\}} \frac{1}{\omega^4},
 & \;\;\;\;\;\; &
 g_3 = 140 \sum\limits_{\omega \in \Lambda \backslash \{0\}} \frac{1}{\omega^6}.
\eq
In both examples the periods can be expressed as integrals involving only algebraic functions.
For the first example we may express the period of the exponential function as
\bq
\label{period_integral_1}
 2 \pi i
 & = &
 4 i \int\limits_0^1 \frac{dt}{\sqrt{1-t^2}}.
\eq
For the second example of Weierstrass's $\wp$-function let us assume that $g_2$ and $g_3$ are two given algebraic numbers.
The periods are expressed as
\bq
\label{period_integral_2}
 \omega_1 = 2 \int\limits_{t_1}^{t_2} \frac{dt}{\sqrt{4t^3-g_2t-g_3}},
 & \;\;\;\;\;\; &
 \omega_2 = 2 \int\limits_{t_3}^{t_2} \frac{dt}{\sqrt{4t^3-g_2t-g_3}},
\eq
where $t_1$, $t_2$ and $t_3$ are the roots of the cubic equation $4t^3-g_2t-g_3=0$.

The representation of the periods of $\exp(z)$ and $\wp(z)$ in the form of eq.~(\ref{period_integral_1}) and eq.~(\ref{period_integral_2})
is the motivation for the following generalisation, due to Kontsevich and Zagier \cite{Kontsevich:2001}:

\index{numerical period}
A numerical period is a complex number whose real and imaginary parts are values
of absolutely convergent integrals of rational functions with rational coefficients,
over domains in $\mathbb{R}^n$ given by polynomial inequalities with rational coefficients.
Domains defined by polynomial inequalities with rational coefficients
are called semi-algebraic sets.

We denote the set of numerical periods by $\mathbb{P}$. 
The numerical periods $\mathbb{P}$ are a countable set of numbers.
We may replace in the above definition every occurrence of ``rational function'' with ``algebraic function''
and every occurrence of ``rational number'' with ``algebraic number'' without changing the set of numbers $\mathbb{P}$.
Then it is clear, that the integrals in eq.~(\ref{period_integral_1}) and eq.~(\ref{period_integral_2}) are numerical periods
in the sense of the above definition, and so is for example $\ln 2$, since
\bq
 \ln 2 & = & \int\limits_1^2 \frac{dt}{t}.
\eq

\section{Elliptic curves}
\label{sec:elliptic}

\vspace{1mm}\noindent
A double lattice $\Lambda$ arises naturally from elliptic curves.
\index{elliptic curve}
Let us consider the elliptic curve
\bq
 E
 & : &
 w^2 - \left(z-z_1\right) \left(z-z_2\right) \left(z-z_3\right) \left(z-z_4\right)
 \; = \; 0,
\eq
where the roots $z_j$ may depend on variables $x=(x_1,...,x_t)$:
\bq
\label{def_roots_generic}
 z_j & = & z_j\left(x\right),
 \;\;\;\;\;\;
 j \in \{1,2,3,4\}.
\eq
We set
\bq
 Z_1 = \left(z_3-z_2\right)\left(z_4-z_1\right),
 \;\;\;
 Z_2 = \left(z_2-z_1\right)\left(z_4-z_3\right),
 \;\;\;
 Z_3 = \left(z_3-z_1\right)\left(z_4-z_2\right).
 \;\;
\eq 
Note that we have $Z_1 + Z_2 = Z_3$.
\index{modulus}
We define the modulus and the complementary modulus of the elliptic curve $E$ by
\bq
 k^2 
 \; = \; 
 \frac{Z_1}{Z_3},
 \;\;\;\;\;\;\;\;\;
 \bar{k}^2 
 \; = \;
 1 - k^2 
 \; = \;
 \frac{Z_2}{Z_3}.
\eq
Note that there are six possibilities of defining $k^2$.
Our standard choice for the periods $\psi_1, \psi_2$ is
\bq
\label{def_generic_periods}
 \psi_1 
 \; = \; 
 \frac{4 K\left(k\right)}{Z_3^{\frac{1}{2}}},
 & &
 \psi_2
 \; = \; 
 \frac{4 i K\left(\bar{k}\right)}{Z_3^{\frac{1}{2}}},
\eq
where $K(x)$ denotes the complete elliptic integral of the first kind.
\begin{figure}
\begin{center}
\includegraphics[scale=1.0]{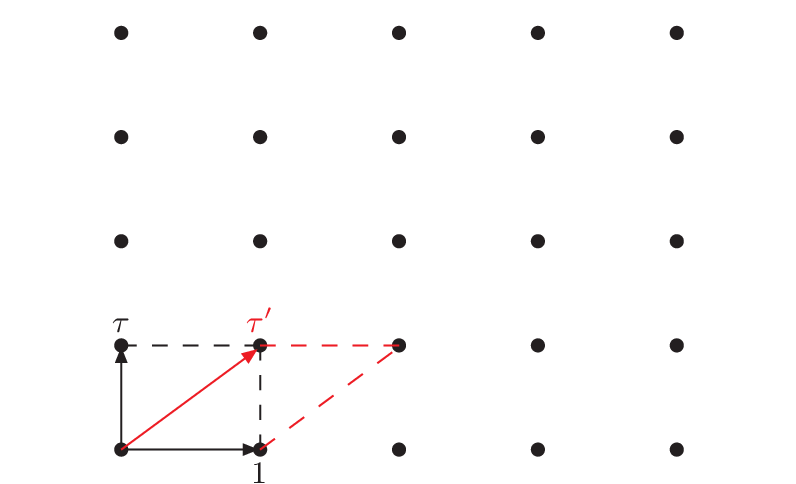}
\end{center}
\caption{The periods $(1,\tau)$ and $(1,\tau')$ generate the same lattice.
}
\label{fig2}
\end{figure}
These two periods generate a lattice $\Lambda=\{n_1 \psi_1 + n_2 \psi_2 \; | \; n_1, n_2 \in {\mathbb Z} \}$.
\index{nome}
We denote the ratio of the two periods and the nome squared by
\bq
\label{def_tau}
 \tau
 \;\; = \;\;
 \frac{\psi_{2}}{\psi_{1}},
 \;\;\;\;\;\;\;\;\;
 q \;\; = \;\; e^{2 i \pi \tau}.
\eq
Let us note that our choice of periods is not unique. Any other choice related to the original one by
\bq
\label{trafo_periods}
 \left( \begin{array}{c}
 \psi_2' \\
 \psi_1' \\
 \end{array} \right)
 \; = \;
 \left( \begin{array}{cc}
 a & b \\
 c & d \\
 \end{array} \right)
 \left( \begin{array}{c}
 \psi_2 \\
 \psi_1 \\
 \end{array} \right),
 \;\;\;\;\;\;\;\;\;\;\;\;
 \left( \begin{array}{cc}
 a & b \\
 c & d \\
 \end{array} \right)
 & \in & \mathrm{SL}\left(2,{\mathbb Z}\right)
\eq
generates the same lattice $\Lambda$.
This is shown in fig.~\ref{fig2}.
\index{modular transformation}
In terms of $\tau$ and $\tau'=\psi_2'/\psi_1'$ the transformation in eq.~(\ref{trafo_periods}) reads
\bq
\label{modular_trafo}
 \tau' & = & \frac{a \tau +b}{c \tau +d}
\eq
and equals a M\"obius transformation.
In this talk we are in particular interested in the situation, where the roots $z_j$ in eq.~(\ref{def_roots_generic}) depend
only on a single variable $x$.
In this case we may exchange the variable $x$ for the variable $\tau$ and study our problem as a function of $\tau$.

\section{Modular forms}
\label{sec:modular}

\vspace{1mm}\noindent
Let us now consider functions of $\tau$.
We are interested in functions with ``nice'' properties under transformations of the form as in eq.~(\ref{modular_trafo}).
We denote by ${\mathbb H} = \lbrace \; \tau \in \mathbb{C} \; | \; \text{Im}(\tau) > 0 \; \rbrace$ the complex upper half plane and by $\overline{\mathbb{H}}$ the extended upper half plane
\bq
 \overline{{\mathbb H}} 
 & = & 
 {\mathbb H} \cup \lbrace \infty \rbrace \cup {\mathbb Q}.
\eq
\index{modular forms}
A meromorphic function $f: \mathbb{H} \rightarrow \mathbb{C}$ is a 
modular form of modular weight $k$ for $\mathrm{SL}\left(2,{\mathbb Z}\right)$ if
\begin{description}
\item{(i)} $f$ transforms under M\"obius transformations as
\bq
\label{weakly_modular}
 f\left( \dfrac{a\tau+b}{c\tau+d} \right) = (c\tau+d)^k \cdot f(\tau) 
 \qquad \text{for} \;\; \left( \begin{array}{cc}
a & b \\ 
c & d
\end{array} \right) \in \mathrm{SL}\left(2,{\mathbb Z}\right)
\eq
\item{(ii)} $f$ is holomorphic on $\mathbb{H}$,
\item{(iii)} $f$ is holomorphic at $\infty$.
\end{description}
We may also look at subgroups of $\mathrm{SL}\left(2,{\mathbb Z}\right)$.
\index{congruence subgroups}
The standard congruence subgroups are defined by
\bq
 \Gamma_0(N) 
 & = &
 \left\{ \left( \begin{array}{cc}
  a & b \\ 
  c & d
 \end{array}  \right) \in \mathrm{SL}\left(2,{\mathbb Z}\right): c \equiv 0\ \text{mod}\ N \right\}, 
 \nonumber \\
 \Gamma_1(N) 
 & = & 
 \left\{ \left( \begin{array}{cc}
  a & b \\ 
  c & d
 \end{array}  \right) \in \mathrm{SL}\left(2,{\mathbb Z}\right): a,d \equiv 1\ \text{mod}\ N, \; c \equiv 0\ \text{mod}\ N  \right\},
 \nonumber \\
 \Gamma(N) 
 & = &
 \left\{ \left( \begin{array}{cc}
  a & b \\ 
  c & d
 \end{array}  \right) \in \mathrm{SL}\left(2,{\mathbb Z}\right): a,d \equiv 1\ \text{mod}\ N, \; b,c \equiv 0\ \text{mod}\ N \right\}.
\eq
Let us also introduce the following notation:
For an integer $k$ and a matrix $\gamma \in \mathrm{SL}\left(2,{\mathbb Z}\right)$ we define 
$f \slashoperator{\gamma}{k}$ by
\bq
(f \slashoperator{\gamma}{k})(\tau) = (c\tau+d)^{-k} \cdot f(\gamma(\tau)).
\eq
With this definition we may re-write the condition (i) in eq.~(\ref{weakly_modular})
as 
\bq
 f \slashoperator{\gamma}{k} = f
 \qquad \text{for all} \;\; \gamma \in \mathrm{SL}\left(2,{\mathbb Z}\right).
\eq
We may now define modular forms for a congruence subgroup $\Gamma$ of $\mathrm{SL}\left(2,{\mathbb Z}\right)$.
A meromorphic function $f: \mathbb{H} \rightarrow \mathbb{C}$ is a 
modular form of modular weight $k$ for $\Gamma$ if
\begin{description}
\item{(i)} $f$ transforms as
\bq
 f \slashoperator{\gamma}{k} = f
 \qquad \text{for all} \;\; \gamma \in \Gamma.
\eq
\item{(ii)} $f$ is holomorphic on $\mathbb{H}$,
\item{(iii)} $f \slashoperator{\alpha}{k}$ is holomorphic at $\infty$ for all $\alpha \in \mathrm{SL}\left(2,{\mathbb Z}\right)$.
\end{description}
For each congruence subgroup $\Gamma$ of $\mathrm{SL}\left(2,{\mathbb Z}\right)$ there is a smallest positive integer
$N$, such that $\Gamma(N) \subseteq \Gamma$.
The integer $N$ is called the level of $\Gamma$.
A modular form $f$ for the congruence subgroup $\Gamma$ of level $N$ has the Fourier expansion
\bq
 f(\tau) 
 & = &
 \sum\limits_{n=0}^{\infty} a_n q^n_N \qquad \text{with} \qquad q_N = e^{2\pi i \tau/N}.
\eq
$f$ is called a cusp form, if $a_0=0$ in the Fourier expansion of $f \slashoperator{\alpha}{k}$ for all
$\alpha \in \mathrm{SL}\left(2,{\mathbb Z}\right)$.

\section{Iterated integrals}
\label{sec:iter_int}

\vspace{1mm}\noindent
\index{iterated integrals}
We review Chen's definition of iterated integrals \cite{Chen}:
Let $M$ be a $t$-dimensional manifold and
\bq
 \gamma & : & \left[0,1\right] \rightarrow M
\eq
a path with start point ${x}_i=\gamma(0)$ and end point ${x}_f=\gamma(1)$.
Suppose further that $\omega_1$, ..., $\omega_k$ are differential $1$-forms on $M$. Let us write
\bq
 f_j\left(\lambda\right) d\lambda & = & \gamma^\ast \omega_j
\eq
for the pull-backs to the interval $[0,1]$.
For $\lambda \in [0,1]$ the $k$-fold iterated integral
of $\omega_1$, ..., $\omega_k$ along the path $\gamma$ is defined by
\bq
 I_{\gamma}\left(\omega_1,...,\omega_k;\lambda\right)
 & = &
 \int\limits_0^{\lambda} d\lambda_1 f_1\left(\lambda_1\right)
 \int\limits_0^{\lambda_1} d\lambda_2 f_2\left(\lambda_2\right)
 ...
 \int\limits_0^{\lambda_{k-1}} d\lambda_k f_k\left(\lambda_k\right).
\eq
We define the $0$-fold iterated integral to be
\bq
 I_{\gamma}\left(;\lambda\right)
 & = &
 1.
\eq
We have
\bq
 \frac{d}{d\lambda}
 I_{\gamma}\left(\omega_1,\omega_2,...,\omega_k;\lambda\right)
 & = &
 f_1\left(\lambda\right) \;
 I_{\gamma}\left(\omega_2,...,\omega_k;\lambda\right).
\eq
Let us now discuss two special cases: Multiple polylogarithms and iterated integrals of modular forms.
\index{multiple polylogarithms}
Multiple polylogarithms are iterated integrals, where all differential one-forms are of the form
\bq
 \gamma^\ast \omega_j & = & \frac{d\lambda}{\lambda-z_j}.
\eq
For $z_w \neq 0$ they are defined by \cite{Goncharov_no_note,Goncharov:2001,Borwein,Moch:2001zr,Vollinga:2004sn}
\bq
 \label{Gfuncdef}
 G(z_1,...,z_w;y)
 & = &
 \int\limits_0^y \frac{dy_1}{y_1-z_1}
 \int\limits_0^{y_1} \frac{dy_2}{y_2-z_2} ...
 \int\limits_0^{y_{w-1}} \frac{dy_w}{y_w-z_w}.
\eq
The number $w$ is referred to as the weight of the multiple polylogarithm
or the depth of the integral representation.
Let us introduce the short-hand notation
\bq
\label{Gshorthand}
 G_{m_1,...,m_k}(z_1,...,z_k;y)
 & = &
 G(\underbrace{0,...,0}_{m_1-1},z_1,...,z_{k-1},\underbrace{0...,0}_{m_k-1},z_k;y),
\eq
where all $z_j$ for $j=1,...,k$ are assumed to be non-zero.
This allows us to relate the integral representation of the multiple polylogarithms 
to the sum representation of the multiple polylogarithms.
The sum representation is defined by
\bq 
\label{def_multiple_polylogs_sum}
 \mathrm{Li}_{m_1,...,m_k}(x_1,...,x_k)
  & = & \sum\limits_{n_1>n_2>\ldots>n_k>0}^\infty
     \frac{x_1^{n_1}}{{n_1}^{m_1}}\ldots \frac{x_k^{n_k}}{{n_k}^{m_k}}.
\eq
The number $k$ is referred to as the depth of the sum representation of the multiple polylogarithm,
the weight is now given by $m_1+m_2+...m_k$.
The relations between the two representations are given by
\bq
\label{Gintrepdef}
 \mathrm{Li}_{m_1,...,m_k}(x_1,...,x_k)
 & = & 
 (-1)^k G_{m_1,...,m_k}\left( \frac{1}{x_1}, \frac{1}{x_1 x_2}, ..., \frac{1}{x_1...x_k};1 \right),
 \nonumber \\
 G_{m_1,...,m_k}(z_1,...,z_k;y) 
 & = & 
 (-1)^k \; \mathrm{Li}_{m_1,...,m_k}\left(\frac{y}{z_1}, \frac{z_1}{z_2}, ..., \frac{z_{k-1}}{z_k}\right).
\eq
If one further sets $g(z;y) = 1/(y-z)$, then one has
\bq
\label{diff_Glog}
 \frac{d}{dy} G(z_1,...,z_w;y) 
 & = & 
 g(z_1;y) G(z_2,...,z_w;y)
\eq
and
\bq
\label{Grecursive}
 G(z_1,z_2,...,z_w;y) & = & \int\limits_0^y dy_1 \; g(z_1;y_1) G(z_2,...,z_w;y_1).
\eq
One can slightly enlarge the set of multiple polylogarithms and define $G(0,...,0;y)$ with $w$ zeros for $z_1$ to $z_w$ to be
\bq
\label{trailingzeros}
 G(0,...,0;y) & = & \frac{1}{w!} \left( \ln y \right)^w.
\eq
This permits us to allow trailing zeros in the sequence $(z_1,...,z_w)$ 
by defining the function $G$ with trailing zeros via eq.~(\ref{Grecursive}) and eq.~(\ref{trailingzeros}).

Our second example are iterated integrals of modular forms.
\index{iterated integrals of modular forms}
Let $f_1(\tau)$, $f_2(\tau)$, ..., $f_k(\tau)$ be modular forms of a congruence subgroup.
Let us further assume that $f_k(\tau)$ vanishes at the cusp $\tau=i\infty$.
For iterated integrals of modular forms we set
\bq
 \omega_j & = & 
 2 \pi i \;\; f_j\left(\tau\right) \; d\tau.
\eq
Thus the $k$-fold iterated integral of modular forms is given by
\bq
 \left(2 \pi i \right)^k
 \int\limits_{i \infty}^{\tau} d\tau_1
 \;
 f_1\left(\tau_1\right)
 \int\limits_{i \infty}^{\tau_1} d\tau_2
 \;
 f_2\left(\tau_2\right)
 ...
 \int\limits_{i \infty}^{\tau_{k-1}} d\tau_k
 \;
 f_k\left(\tau_k\right).
\eq
The case where $f_k(\tau)$ does not vanishes at the cusp $\tau=i\infty$ is discussed in \cite{Adams:2017ejb,Brown:2014aa} and is similar to trailing zeros in the case of multiple polylogarithms.

\section{Precision calculations}
\label{sec:precision}

\vspace{1mm}\noindent
\index{perturbation theory}
Due to the smallness of all coupling constants $g$, 
we may compute at high energies an infrared-safe observable (for example the cross section $\sigma$ for a particular process)
reliable in perturbation theory:
\bq
 \sigma & = & 
 \left( \frac{g}{4\pi} \right)^4 \sigma_{LO}
 + \left( \frac{g}{4\pi} \right)^6 \sigma_{NLO}
 + \left( \frac{g}{4\pi} \right)^8 \sigma_{NNLO}
 + ...
\eq
The cross section is related to the square of the scattering amplitude
\bq
 \sigma \sim \left| {\mathcal A} \right|^2,
\eq
and the perturbative expansion of the cross section follows from the 
perturbative expansion of the amplitude
\bq
 {\mathcal A}
 & = &
 g^2
 {\mathcal A}^{(0)}
 +
 g^4
 {\mathcal A}^{(1)}
 +
 g^6
 {\mathcal A}^{(2)}
 +
 ...,
\eq
where ${\mathcal A}^{(l)}$ contains $l$ loops.
The computation of the tree amplitude ${\mathcal A}^{(0)}$ poses no conceptional problem.
For loop amplitudes we have to calculate Feynman integrals.
Let us write
\bq
 {\mathcal A}^{(l)}
 & = &
 \sum\limits_j c_j I_j,
\eq
where the $I_j$'s are Feynman integrals and the $c_j$'s are coefficients, whose computation is tree-like.
Without loss of generality we may 
take the set of Feynman integrals $\{I_1,I_2,...\}$ to consist of scalar integrals \cite{Tarasov:1996br,Tarasov:1997kx}.
Let us now look closer on the Feynman integrals.
\index{Feynman integrals}
A Feynman graph $G$ with $n$ external lines, $r$ internal lines and $l$ loops corresponds 
(up to prefactors)
in $D$ space-time dimensions to
the family of Feynman integrals,
indexed by the powers of the propagators $\nu_j$
\bq
\label{def_feynman_integral_prop}
I^G_{\nu_1 \nu_2 ... \nu_r}  & = &
 \frac{\prod\limits_{j=1}^{r}\Gamma(\nu_j)}{\Gamma(\nu-lD/2)}
 \;
 \left( \mu^2 \right)^{\nu-l D/2}
 \;\;
 \int \prod\limits_{s=1}^{l} \frac{d^Dk_s}{i\pi^{\frac{D}{2}}}
 \;\;
 \prod\limits_{j=1}^{r} \frac{1}{(-q_j^2+m_j^2)^{\nu_j}},
\eq
with $\nu=\nu_1+...+\nu_r$.
The momenta flowing through the internal lines can be expressed through the 
independent loop momenta $k_1$, ..., $k_l$ and the external momenta $p_1$, ..., $p_n$ as 
\bq
 q_i & = & \sum\limits_{j=1}^l \lambda_{ij} k_j + \sum\limits_{j=1}^n \sigma_{ij} p_j,
 \;\;\;\;\;\;\;\;\; 
 \lambda_{ij}, \sigma_{ij} \in \{-1,0,1\}.
\eq
\index{graph polynomials}
After Feynman parametrisation we obtain
\bq
\label{feynman_integral}
 I^G_{\nu_1 \nu_2 ... \nu_r}  
 & = &
 \int\limits_{\Delta}  \Omega
 \left( \prod\limits_{j=1}^r x_j^{\nu_j-1} \right)
 \frac{{\mathcal U}^{\nu-(l+1) D/2}}{{\mathcal F}^{\nu-l D/2}}.
\eq
The prefactors in the definition of the Feynman integral in eq.~(\ref{def_feynman_integral_prop}) are chosen such that
after Feynman parametrisation we obtain an expression without prefactors, as can be seen from eq.~(\ref{feynman_integral}).
In eq.~(\ref{feynman_integral}) the integration is over
\bq
 \Delta & = & \left\{ \left[ x_1 : x_2 : ... : x_r \right] \in {\mathbb P}^{r-1} | x_i \ge 0 \right\}.
\eq
Here, ${\mathbb P}^{r-1}$ denotes the real projective space with $r-1$ dimensions.
$\Omega$ is a differential $(r-1)$-form given by
\bq
 \Omega 
 & = & 
 \sum\limits_{j=1}^r (-1)^{j-1}
  \; x_j \; dx_1 \wedge ... \wedge \widehat{dx_j} \wedge ... \wedge dx_r,
\eq
where the hat indicates that the corresponding term is omitted.
The functions ${\mathcal U}$ and ${\mathcal F}$ are obtained from first writing
\bq
 \sum\limits_{j=1}^{r} x_{j} (-q_j^2+m_j^2)
 & = & 
 - \sum\limits_{a=1}^{l} \sum\limits_{b=1}^{l} k_a M_{ab} k_b + \sum\limits_{a=1}^{l} 2 k_a \cdot Q_a - J,
\eq
where $M$ is a $l \times l$ matrix with scalar entries and $Q$ is a $l$-vector
with $D$-vectors as entries.
We then have
\bq
\label{definition_U_and_F}
 {\mathcal U} = \mbox{det}(M),
 & &
 {\mathcal F} = \mbox{det}(M) \left( - J + Q M^{-1} Q \right)/\mu^2.
\eq 
${\mathcal U}$ and ${\mathcal F}$ are the first and second graph polynomial of the Feynman graph $G$ \cite{Bogner:2010kv}.

The Feynman integral defined in eq.~(\ref{feynman_integral}) has an expansion as a Laurent series 
in the parameter $\eps=(4-D)/2$ of dimensional regularisation:
\bq
\label{epsilon_expansion}
 I^G_{\nu_1 \nu_2 ... \nu_r} & = &
 \sum\limits_{j=j_{\mathrm{min}}}^\infty f_j \eps^j.
\eq
The coefficients $f_j$ are in general functions of the Lorentz invariants
\bq
 s_J & = & \left( \sum\limits_{j \in J} p_j \right)^2,
\eq
where the sum runs over a subset $J$ of the external momenta, the internal masses $m_i$ and the scale $\mu$.
We are interested in the question, to which class of functions the coefficients $f_j$ belong.
Let us first consider the situation, where we keep all Lorentz invariants, all masses and the scale fixed.
Suppose that
(i) all kinematical invariants $s_J$ are negative or zero, 
(ii) all masses $m_i$ and $\mu$ are positive or zero ($\mu \neq 0$) and
(iii) all ratios of invariants and masses are rational,
then it can be shown that all coefficients $f_j$
in eq.~(\ref{epsilon_expansion}) are numerical periods \cite{Bogner:2007mn}.

Let us now return to the original problem and view the coefficients $f_j$ as functions of the Lorentz invariants $s_J$,
the internal masses $m_i$ and the scale $\mu$.
Let us consider a family of Feynman integrals $I^G_{\nu_1 \nu_2 ... \nu_r}$, including all its sub-topologies.
A sub-topology $G'$ is obtained by pinching in the graph $G$ one or several internal lines.
In the Feynman integral the corresponding propagators are then absent and the associated exponents $\nu_j$ are zero.
\begin{figure}
\begin{center}
\includegraphics[scale=0.85]{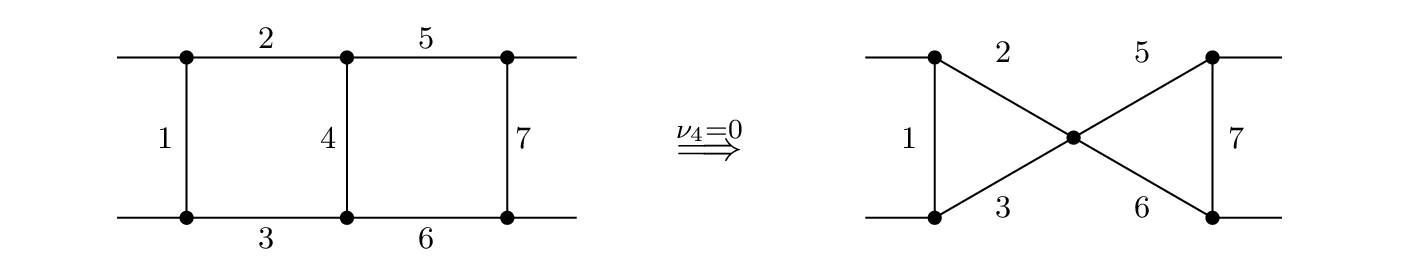}
\end{center}
\caption{If for some exponent we have $\nu_j=0$, the corresponding propagator is absent and the 
topology simplifies.
}
\label{fig1}
\end{figure}
This is shown in fig.~\ref{fig1}.
Integration-by-parts identities \cite{Tkachov:1981wb,Chetyrkin:1981qh} 
allow us to express the Feynman integrals from the family $I^G_{\nu_1 \nu_2 ... \nu_r}$
as a linear combination of a few master integrals, which we denote by $I = \{I_1,...,I_N\}$.
Let us further denote by $x = \left(x_1,...,x_t\right)$ the vector of kinematic variables the master integrals depend on.
\label{differential equations for Feynman integrals}
The method of differential equations \cite{Kotikov:1990kg,Kotikov:1991pm,Remiddi:1997ny,Gehrmann:1999as,Argeri:2007up,MullerStach:2012mp,Henn:2013pwa,Henn:2014qga,Ablinger:2015tua,Bosma:2017hrk} is a powerful tool 
to find the functions $f_j$ in eq.~(\ref{epsilon_expansion}).
Let $x_k$ be a kinematic variable.
Carrying out the derivative $\partial I_i/\partial x_k$ 
under the integral sign and using integration-by-parts identities allows us to express the 
derivative as a linear combination of the master integrals:
\bq
 \frac{\partial}{\partial x_k} I_i
 +
 \sum\limits_{j=1}^N a_{ij} I_j
 & = &
 0.
\eq
\index{Gau{\ss}-Manin connection}
Repeating the above procedure for every master integral and every kinematic variable we 
obtain a system of differential equations of Fuchsian type
\bq
\label{diff_eq}
 \left( d + A \right) I 
 & = & 0,
\eq
where $A$ is a matrix-valued one-form
\bq
 A & = & 
 \sum\limits_{i=1}^t A_i dx_i.
\eq
The matrix-valued one-form $A$ satisfies the integrability condition $dA + A \wedge A = 0$.

Geometrically we have a vector bundle with a flat connection: The base space is parametrised by the coordinates
$x=(x_1,...,x_t)$, the fibre is a $N$-dimensional vector space with basis $I=(I_1,...,I_N)$, the flat connection
is given by $A$ and called the Gau{\ss}-Manin connection.

Suppose $A$ is of the form
\bq
\label{eps_form_polylogs}
 A 
 & = &
 \eps \sum\limits_j \; C_j \; d \ln p_j\left(x\right),
\eq
where all $\eps$-dependence is in the prefactor, the $C_j$'s are matrices with constant entries
and the $p_j(x)$'s are polynomials in the external variables $x$, then the system of differential
equations is easily solved in terms of multiple polylogarithms \cite{Henn:2013pwa}.

In this talk we consider the situation, where the master integrals depend only on a single variable $\tau$ and
the connection one-form $A$ is of the form
\bq
\label{eps_form_modular_forms}
 A
 & = &
 \eps \sum\limits_j \; F_j \; \left(2 \pi i\right) \; d\tau,
\eq
where as before all $\eps$-dependence is in the prefactor and the $F_j$'s are matrices, whose entries
are modular forms.
In this case the system of differential equations is easily solved in terms of iterated integrals
of modular forms.

A system of differential equations, where the only $\eps$-dependence is in a prefactor like in eq.~(\ref{eps_form_polylogs}) or eq.~(\ref{eps_form_modular_forms}) is said to be in $\eps$-form.
Clearly, it is advantageous to have the system in $\eps$-form.
There are two operations at our disposal to transform a system of differential equations, which follow
from the geometric picture described above:
We may change the variables in the base manifold and/or we may change the basis of the vectorspace in the fibre.
A change of variables in the base manifold introduces a Jacobian: If $\tau'=\gamma(\tau)$ (for simplicity we consider the case where the base manifold is one-dimensional)
we have
\bq
 A' & = & A \; \frac{\partial \tau'}{\partial \tau}.
\eq
A change of the basis of the vectorspace in the fibre
\bq
 I' & = & U I
\eq
transforms the connection into
\bq
 A' & = & U A U^{-1} + U d U^{-1}.
\eq

\section{Picard-Fuchs operators}
\label{sec:picard_fuchs}

\vspace{1mm}\noindent
\index{Picard-Fuchs operators}
An extremely helpful tool for Feynman integral computations within the approach based on differential equations 
are the factorisation properties of Picard-Fuchs operators \cite{Adams:2017tga}.
Let us consider an (unknown) function $f(\lambda)$ of a single variable $\lambda$, which obeys a (known)
homogeneous differential equation
of order $r$
\bq
  \sum\limits_{j=0}^r p_j(\lambda) \frac{d^j}{d\lambda^j} f(\lambda) & = & 0,
\eq
where the $p_j$'s are polynomials in $\lambda$, such that the differential equation is of Fuchsian type.
We call the differential operator
\bq
 L & = & \sum\limits_{j=0}^r p_j(\lambda) \frac{d^j}{d\lambda^j}
\eq
a Picard-Fuchs operator.
Suppose that this operator factorises into linear factors:
\bq
 L
 & = & 
 \left( a_r(\lambda) \frac{d}{d\lambda} + b_r(\lambda) \right)
 ...
 \left( a_2(\lambda) \frac{d}{d\lambda} + b_2(\lambda) \right)
 \left( a_1(\lambda) \frac{d}{d\lambda} + b_1(\lambda) \right).
 \;\;\;
\eq
Such a differential equation is easily solved.
Let us denote the homogeneous solution of the $j$-th factor by
\bq
 \psi_j(\lambda) & = & \exp\left(- \int\limits_0^\lambda d\kappa \; \frac{b_j(\kappa)}{a_j(\kappa)} \right).
\eq
Then the full solution is given by iterated integrals as
\bq
 f(\lambda)  & = &
 C_1 \psi_1(\lambda) 
 + C_2 \psi_1(\lambda) \int\limits_0^\lambda d\lambda_1 \frac{\psi_2(\lambda_1)}{a_1(\lambda_1) \psi_1(\lambda_1)} 
 \nonumber \\
 & & 
 + C_3 \psi_1(\lambda) \int\limits_0^\lambda d\lambda_1 \frac{\psi_2(\lambda_1) }{a_1(\lambda_1) \psi_1(\lambda_1)} \int\limits_0^{\lambda_1} d\lambda_2 \frac{\psi_3(\lambda_2)}{a_2(\lambda_2) \psi_2(\lambda_2)} 
 + ...
\eq
From eq.~(\ref{diff_Glog}) we see that multiple polylogarithms are of this form, i.e. have Picard-Fuchs operators, which factorise into linear
factors.

The next more complicated situation is the case, where the Picard-Fuchs operator contains one irreducible second-order
differential operator
\bq
 a_j(\lambda) \frac{d^2}{d\lambda^2} + b_j(\lambda) \frac{d}{d\lambda} + c_j(\lambda).
\eq
As an example consider the differential equation
\bq
 \left[ \lambda \left(1-\lambda^2\right) \frac{d^2}{d\lambda^2} + \left(1-3\lambda^2\right) \frac{d}{d\lambda} - \lambda \right] f(\lambda) & = & 0
\eq
This second-order differential operator 
is irreducible.
The solutions of the differential equation are $K(\lambda)$ and $K(\sqrt{1-\lambda^2})$,
where $K(\lambda)$ is the complete elliptic integral of the first kind:
\bq
 K(\lambda)
 & = &
 \int\limits_0^1 \frac{dx}{\sqrt{\left(1-x^2\right)\left(1-\lambda^2x^2\right)}}.
\eq
Let us now return to a system of differential equations as in eq.~(\ref{diff_eq}).
In general, such a system may depend on several kinematic variables $x=(x_1,...,x_t)$.
We may reduce a multi-scale system to a single-scale system by
setting $x_i\left(\lambda\right) = \alpha_i \lambda$ with 
$\alpha=[\alpha_1:...:\alpha_t] \in {\mathbb C} {\mathbb P}^{t-1}$
and by viewing the master integrals as functions of $\lambda$.
For the derivative with respect to $\lambda$ we have
\bq
 \frac{d}{d\lambda} I
 & = &
 B I,
 \;\;\;\;\;\;
 B \; = \;
 \sum\limits_{i=1}^t \alpha_i A_i.
\eq
In addition we may assume that the $\eps$-dependence of the matrices $A$ and $B$ is polynomial, if this is not the case, a rescaling of the master integrals
with $\eps$-dependent prefactors will achieve this situation.
Let us write
\bq
 \;\;\;\;\;\;
 B \; = \; B^{(0)} + \sum\limits_{j>0} \eps^j B^{(j)}.
\eq
A system of ordinary first-order differential equations is easily 
converted to a higher-order differential equation for a single master integral.
We may work modulo sub-topologies, therefore the order of the differential equation is given by the number $N_s$
of master integrals in this sector.
In order to find the required transformation we work in addition modulo $\eps$-corrections, i.e. 
we focus on $B^{(0)}$.
Let $I$ be one of the master integrals $\{I_1,...,I_{N_s}\}$.
We determine the largest number $r$, such that the matrix which expresses 
$I$, $(d/d\lambda)I$, ..., $(d/d\lambda)^{r-1}I$ in terms of the original set $\{I_1,...,I_{N_s}\}$ has full rank.
It follows that $(d/d\lambda)^rI$ can be written as a linear combination of $I, ..., (d/d\lambda)^{r-1}I$.
This defines the Picard-Fuchs operator $L$ for the master integral $I$ with respect to $\lambda$:
\bq
 L I & = & 0,
 \;\;\;\;\;\;
 L \; = \; \sum\limits_{k=0}^r p_k(\lambda) \frac{d^k}{d\lambda^k}.
\eq
$L$ is easily found by transforming to a basis which contains $I, ..., (d/d\lambda)^{r-1}I$.
\index{factorisation of differential operators}
Although the Picard-Fuchs operator is a differential operator of order $r$, it is very often the
case that this operator factorises.
The factorisation can be obtained with standard algorithms \cite{vanHoeij:1997}.
Let us write for the factorisation into irreducible factors
\bq
 L
 & = &
 L_{1} L_{2} ... L_{s}, 
\eq
where the differential operators $L_{i}$ are irreducible. 
Since we started from the $\eps$-independent matrix $B^{(0)}$, the differential operators $L_i$
are $\eps$-independent.

\section{Feynman integrals evaluating to iterated integrals of modular forms}
\label{sec:examples}

\vspace{1mm}\noindent
Let us now consider a few examples. 
\begin{figure}
\begin{center}
\includegraphics[scale=1.0]{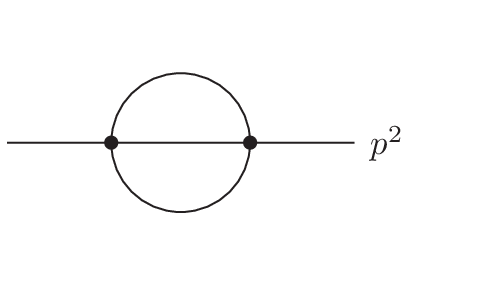}
\includegraphics[scale=1.0]{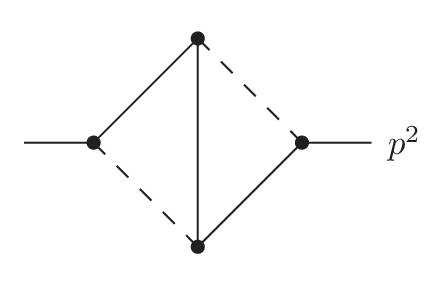}
\includegraphics[scale=0.9]{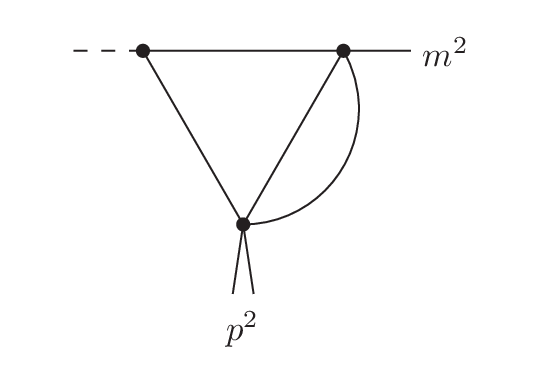}
\includegraphics[scale=0.9]{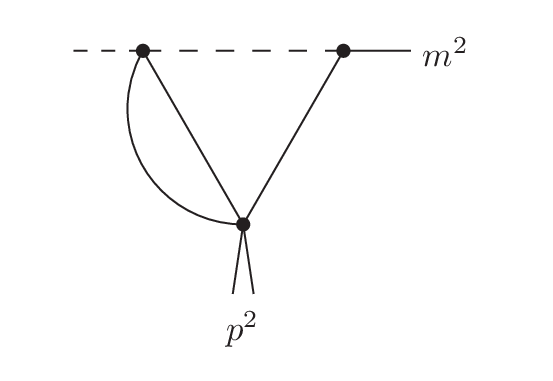}
\end{center}
\caption{
Examples of Feynman integrals evaluating to iterated integrals of modular forms.
Internal solid lines correspond to a propagator with mass $m^2$, internal dashed lines to a massless 
propagator. External dashed lines indicate a light-like external momentum.
}
\label{fig3}
\end{figure}
We consider the Feynman integrals shown in fig.~\ref{fig3}.
These are two-loop two-point or three-point integrals, depending on a single dimensionless variable
\bq
 x & = & \frac{p^2}{m^2}.
\eq
All examples shown in fig.~\ref{fig3} contain the equal-mass sunrise graph as a subtopology
and are -- as we will see -- expressible in terms of iterated integrals of modular forms. 
In order to proceed we would like to
\begin{enumerate}
\item verify that the integrals depend only on a single elliptic curve,
\item identify the elliptic curve,
\item change the variable of the base manifold from $x$ to the modular parameter $\tau$,
\item change the basis of master integrals such that the transformed system of differential equations
is in $\eps$-form.
\end{enumerate}
These steps can be done systematically. Let us start with the first step.
In order to verify that the integrals depend only on a single elliptic curve we construct
for all integrals (including all sub-topologies) the Picard-Fuchs operators as described in the previous section.
We recall that for a specific integral we work modulo sub-topologies and modulo $\eps$-corrections.
We then look at the factorisations of the various Picard-Fuchs operators and verify, that there is only one second-order
irreducible factor. All other factors are first order.
The irreducible second-order differential operator is associated with the sunrise graph.

In the second step we identify the elliptic curve. For the sunrise graph this can be done either from 
the maximal cuts \cite{Baikov:1996iu,Lee:2009dh,Kosower:2011ty,CaronHuot:2012ab,Frellesvig:2017aai,Bosma:2017ens,Harley:2017qut}
or from the Feynman parameter representation. 
The former method generalises easily to more complicated Feynman integrals \cite{Adams:2018bsn,Adams:2018kez} and we discuss
it here.
One finds for the sunrise integral in two space-time dimensions
\bq
\label{maxcut}
 \mathrm{MaxCut}_{\mathcal C} \; I
 & = &
 \frac{u}{\pi^2}
 \int\limits_{\mathcal C} 
 \frac{dz}{z^{\frac{1}{2}} \left(z + 4 \right)^{\frac{1}{2}} \left[z^2 + 2 \left(1+x\right) z + \left(1-x\right)^2 \right]^{\frac{1}{2}}},
\eq
where $u$ is an (irrelevant) phase and ${\mathcal C}$ an integration contour.
The denominator of the integrand defines an elliptic curve, which we denote by $E_x$:
\bq
\label{E_73_maxcut}
 E_x & : &
 w^2 - z
       \left(z + 4 \right) 
       \left[z^2 + 2 \left(1+x\right) z + \left(1-x\right)^2 \right]
 \; = \;
 0.
\eq
We denote the roots of the quartic polynomial in eq.~(\ref{E_73_maxcut}) by
\bq
 z_1 \; = \; -4,
 \;\;\;
 z_2 \; = \; -\left(1+\sqrt{x}\right)^2,
 \;\;\;
 z_3 \; = \; -\left(1-\sqrt{x}\right)^2,
 \;\;\;
 z_4 \; = \; 0.
\eq
We consider a neighbourhood of $x=0$ without the branch cut
of $\sqrt{x}$ along the negative real axis.
The correct physical value is specified by Feynman's $i\delta$-prescription: $x\rightarrow x+i\delta$.
The periods $\psi_1, \psi_2$ and the modular parameter $\tau$ are then defined by
eq.~(\ref{def_generic_periods}) and eq.~(\ref{def_tau}), respectively.

In the third step we change the variable of the base manifold from $x$ to the modular parameter $\tau$.
We recall that $\tau$ as a function of $x$ is given by eq.~(\ref{def_tau}):
\bq
\label{def_tau_2}
 \tau
 & = &
 \frac{\psi_{2}}{\psi_{1}}.
\eq
In a neighbourhood of $x=0$ we may invert eq.~(\ref{def_tau_2}). This gives
\bq
\label{hauptmodul}
 x
 & = &
 9 \frac{\eta\left(6\tau\right)^8 \eta\left(\tau\right)^4}
        {\eta\left(2\tau\right)^8 \eta\left(3\tau\right)^4},
\eq
where $\eta$ denotes Dedekind's eta-function. For the Jacobian we have
\bq
 \frac{d\tau}{dx} & = & \frac{W}{\psi_1^2},
\eq
where the Wronskian $W$ is given by
\bq
\label{Wronskian_relation}
 W
 & = & 
 \psi_{1} \frac{d}{dx} \psi_{2} - \psi_{2} \frac{d}{dx} \psi_{1}
 \;\; = \;\;
 - \frac{6 \pi i}{x\left(x-1\right)\left(x-9\right)}.
\eq
In the fourth step we change the basis of master integrals such that the transformed system of differential equations
is in $\eps$-form.
The essential new ingredient is the appropriate definition of the master integrals corresponding to the
second-order irreducible differential operator.
We need two master integrals for this case.
The first master integral may be taken as the sunrise integral in $D=2-2\eps$ space-time dimensions divided by the 
$\eps^0$-term of its maximal cut.
This is familiar from the case of Feynman integrals, which evaluate to multiple polylogarithms.
The difference lies in the fact, that for Feynman integrals, which evaluate to multiple polylogarithms, the maximal
cut is an algebraic function, while in the case of the sunrise integral it is given by a complete elliptic integral.
We thus set
\bq
 I_1 
 & = &
  \eps^2 \frac{\pi}{\psi_1} S_{111}\left(2-2\eps,x\right),
\eq
where $S_{111}(2-2\eps,x)$ denotes the sunrise integral in $D=2-2\eps$ space-time dimensions with $\nu_1=\nu_2=\nu_3=1$.
Let us turn to the second master integral: It is well-known in mathematics, that the first cohomology group for
a family of elliptic curves $E_x$, parametrised by $x$, is generated by the holomorphic one form $dz/w$ and its $x$-derivative.
This motivates an ansatz, consisting of $I_1$ and its $\tau$-derivative.
One finds for the second master integral in the elliptic sector
\bq
 I_2
 & = &
 \frac{1}{\eps} \frac{1}{2\pi i} \frac{d}{d\tau} I_1
 +
 \frac{1}{24} \left(3x^2-10x-9\right) \frac{\psi_1^2}{\pi^2} I_1.
\eq
The full set of master integrals is completed by transforming in addition the master integrals in the non-elliptic
sectors.
The entries on the diagonal of the transformation matrix for the non-elliptic sectors can be read off
from the linear factors appearing in the factorisation of the Picard-Fuchs operators \cite{Adams:2017tga}.
The non-diagonal entries are obtained from an ansatz along the lines of \cite{Meyer:2016slj,Meyer:2017joq}.

Let us look at a specific example. We denote the two-loop tadpole integral by
\bq
 I_0
 & = &
 4 \eps^2 S_{110}\left(2-2\eps,x\right).
\eq
Then we have for $I=(I_0,I_1,I_2)$
\bq
\label{res_eps_form}
 \frac{1}{2\pi i} \frac{d}{d\tau} I
 & = &
 \eps \; A \; I,
\eq
where the matrix $A$ is $\eps$-independent and is given by
\bq
\label{res_A}
 A & = & 
 \left( \begin{array}{rrr}
 0 & 0 & 0 \\
 0 & -f_2 & 1 \\
 \frac{1}{4} f_3 & f_4 & -f_2 \\
 \end{array} \right).
\eq
The entries of $A$ are given by
\bq
\label{integration_kernels}
 f_{2} 
 & = &
 \frac{1}{2 i \pi} \frac{\psi_{1}^2}{W} \frac{\left(3x^2-10 x - 9 \right)}{2 x \left(x-1\right) \left(x-9\right)},
 \nonumber \\
 f_{3} 
 & = &
 \frac{\psi_{1}^3}{4 \pi W^2}
 \;
 \frac{6}{x \left(x-1\right)\left(x-9\right)},
 \nonumber \\
 f_{4}
 & = &
 \frac{1}{576}
 \frac{\psi_{1}^4}{\pi^4} \; \left(x+3\right)^4.
\eq
One checks that $f_2$, $f_3$ and $f_4$ are modular forms of $\Gamma_1(6)$ of modular weight $2$, $3$ and $4$, respectively.
We introduce a basis $\{e_1,e_2\}$ for the modular forms of modular weight $1$ 
for the Eisenstein subspace ${\mathcal E}_1(\Gamma_1(6))$:
\bq
 e_1 \; = \; E_1\left(\tau;\chi_0,\chi_1\right),
 & &
 e_2 \; = \; E_1\left(2\tau;\chi_0,\chi_1\right),
\eq
where $E_1(\tau,\chi_0,\chi_1)$ and $E_1(2\tau,\chi_0,\chi_1)$
are generalised Eisenstein series \cite{Stein} and
$\chi_0$ and $\chi_1$ denote primitive Dirichlet characters with conductors $1$ and $3$, respectively.
The integration kernels may be expressed as polynomials in $e_1$ and $e_2$:
\bq
 f_2
 & = &
 -6 \left( e_1^2 + 6 e_1 e_2 - 4 e_2^2 \right),
 \nonumber \\
 f_3
 & = &
 36 \sqrt{3} \left( e_1^3 - e_1^2 e_2 - 4 e_1 e_2^2 + 4e_2^3 \right),
 \nonumber \\
 f_4 
 & = &
 324 e_1^4.
\eq
The solution for these Feynman integrals in terms of iterated integrals of modular forms follows now directly from the
differential equation~(\ref{res_eps_form}).
The $q$-expansion of the iterated integrals provides an efficient method for the numerical evaluation \cite{Bogner:2017vim,Bognerandmore}.

Let us close this paragraph with the observation that the integration kernels
\bq
 \omega_0 \; = \; \frac{dx}{x},
 \;\;\;\;\;\;
 \omega_0 \; = \; \frac{dx}{x-1}
\eq
may also be expressed as modular forms:
\bq
 \omega_0 \; = \; g_{2,0} \; 2 \pi i \; d\tau,
 \;\;\;\;\;\;
 \omega_0 \; = \; g_{2,1} \; 2 \pi i \; d\tau.
\eq
The modular forms $g_{2,0}$ and $g_{2,1}$, both of modular weight $2$, are given by
\bq
 g_{2,0} 
 & = &
 \frac{1}{2 i \pi} \frac{\psi_{1}^2}{W} \frac{1}{x}
 \; =  \;
 - 12 \left( e_1^2 - 4 e_2^2 \right),
 \nonumber \\
 g_{2,1} 
 & = &
 \frac{1}{2 i \pi} \frac{\psi_{1}^2}{W} \frac{1}{x-1}
 \; = \;
 - 18 \left( e_1^2 + e_1 e_2 - 2 e_2^2 \right).
\eq
This shows that the harmonic polylogarithms \cite{Vermaseren:1998uu,Remiddi:1999ew} in the letters $0$ and $1$ 
are a subset of the iterated
integrals of modular forms discussed in this talk.

\section{Conclusions}
\label{sec:conclusions}

\vspace{1mm}\noindent
In this talk we considered a class of Feynman integrals, which evaluate to iterated integrals
of modular forms.
These Feynman integrals are beyond the class of Feynman integrals, which evaluate to multiple
polylogarithms.
However, several important properties, known from the case of multiple polylogarithms, carry over:
The system of differential equations can be brought into an $\eps$-form, the iterated integrals
satisfy a shuffle algebra and there is an efficient method for the numerical evaluation 
of the iterated integrals of modular forms based on the $q$-expansion.
We considered single-scale integrals. We may view these Feynman integrals, which evaluate
to iterated integrals of modular forms as generalisations of Feynman integrals, which may be expressed
in terms of harmonic polylogarithms in the letters $0$ and $1$.

\vspace*{4mm}
\noindent
{\bf Acknowledgement.} S.W. would like to thank the organisers and KMPB for the organisation of the
inspiring conference.



\begin{thebibliography}{10}

\bibitem{Broadhurst:1993mw}
D.~J. Broadhurst, J.~Fleischer, and O.~Tarasov,
\newblock Z.Phys. {\bf C60}, 287 (1993), arXiv:hep-ph/9304303.

\bibitem{Berends:1993ee}
F.~A. Berends, M.~Buza, M.~B{\"o}hm, and R.~Scharf,
\newblock Z.Phys. {\bf C63}, 227 (1994).

\bibitem{Bauberger:1994nk}
S.~Bauberger, M.~B{\"o}hm, G.~Weiglein, F.~A. Berends, and M.~Buza,
\newblock Nucl.Phys.Proc.Suppl. {\bf 37B}, 95 (1994), arXiv:hep-ph/9406404.

\bibitem{Bauberger:1994by}
S.~Bauberger, F.~A. Berends, M.~B{\"o}hm, and M.~Buza,
\newblock Nucl.Phys. {\bf B434}, 383 (1995), arXiv:hep-ph/9409388.

\bibitem{Bauberger:1994hx}
S.~Bauberger and M.~B{\"o}hm,
\newblock Nucl.Phys. {\bf B445}, 25 (1995), arXiv:hep-ph/9501201.

\bibitem{Caffo:1998du}
M.~Caffo, H.~Czyz, S.~Laporta, and E.~Remiddi,
\newblock Nuovo Cim. {\bf A111}, 365 (1998), arXiv:hep-th/9805118.

\bibitem{Laporta:2004rb}
S.~Laporta and E.~Remiddi,
\newblock Nucl. Phys. {\bf B704}, 349 (2005), hep-ph/0406160.

\bibitem{Kniehl:2005bc}
B.~A. Kniehl, A.~V. Kotikov, A.~Onishchenko, and O.~Veretin,
\newblock Nucl. Phys. {\bf B738}, 306 (2006), arXiv:hep-ph/0510235.

\bibitem{Groote:2005ay}
S.~Groote, J.~G. K{\"o}rner, and A.~A. Pivovarov,
\newblock Annals Phys. {\bf 322}, 2374 (2007), arXiv:hep-ph/0506286.

\bibitem{Groote:2012pa}
S.~Groote, J.~K{\"o}rner, and A.~Pivovarov,
\newblock Eur.Phys.J. {\bf C72}, 2085 (2012), arXiv:1204.0694.

\bibitem{Bailey:2008ib}
D.~H. Bailey, J.~M. Borwein, D.~Broadhurst, and M.~L. Glasser,
\newblock J. Phys. {\bf A41}, 205203 (2008), arXiv:0801.0891.

\bibitem{MullerStach:2011ru}
S.~M{\"u}ller-Stach, S.~Weinzierl, and R.~Zayadeh,
\newblock Commun. Num. Theor. Phys. {\bf 6}, 203 (2012), arXiv:1112.4360.

\bibitem{Adams:2013nia}
L.~Adams, C.~Bogner, and S.~Weinzierl,
\newblock J. Math. Phys. {\bf 54}, 052303 (2013), arXiv:1302.7004.

\bibitem{Bloch:2013tra}
S.~Bloch and P.~Vanhove,
\newblock J. Numb. Theor. {\bf 148}, 328 (2015), arXiv:1309.5865.

\bibitem{Adams:2014vja}
L.~Adams, C.~Bogner, and S.~Weinzierl,
\newblock J. Math. Phys. {\bf 55}, 102301 (2014), arXiv:1405.5640.

\bibitem{Adams:2015gva}
L.~Adams, C.~Bogner, and S.~Weinzierl,
\newblock J. Math. Phys. {\bf 56}, 072303 (2015), arXiv:1504.03255.

\bibitem{Adams:2015ydq}
L.~Adams, C.~Bogner, and S.~Weinzierl,
\newblock J. Math. Phys. {\bf 57}, 032304 (2016), arXiv:1512.05630.

\bibitem{Remiddi:2013joa}
E.~Remiddi and L.~Tancredi,
\newblock Nucl.Phys. {\bf B880}, 343 (2014), arXiv:1311.3342.

\bibitem{Bloch:2016izu}
S.~Bloch, M.~Kerr, and P.~Vanhove,
\newblock Adv. Theor. Math. Phys. {\bf 21}, 1373 (2017), arXiv:1601.08181.

\bibitem{Groote:2018rpb}
S.~Groote and J.~G. K{\"o}rner,
\newblock (2018), arXiv:1804.10570.

\bibitem{Bloch:2014qca}
S.~Bloch, M.~Kerr, and P.~Vanhove,
\newblock Compos. Math. {\bf 151}, 2329 (2015), arXiv:1406.2664.

\bibitem{Remiddi:2016gno}
E.~Remiddi and L.~Tancredi,
\newblock Nucl. Phys. {\bf B907}, 400 (2016), arXiv:1602.01481.

\bibitem{Adams:2016xah}
L.~Adams, C.~Bogner, A.~Schweitzer, and S.~Weinzierl,
\newblock J. Math. Phys. {\bf 57}, 122302 (2016), arXiv:1607.01571.

\bibitem{Adams:2017ejb}
L.~Adams and S.~Weinzierl,
\newblock Commun. Num. Theor. Phys. {\bf 12}, 193 (2018), arXiv:1704.08895.

\bibitem{Bogner:2017vim}
C.~Bogner, A.~Schweitzer, and S.~Weinzierl,
\newblock Nucl. Phys. {\bf B922}, 528 (2017), arXiv:1705.08952.

\bibitem{Adams:2018yfj}
L.~Adams and S.~Weinzierl,
\newblock Phys. Lett. {\bf B781}, 270 (2018), arXiv:1802.05020.

\bibitem{Adams:2018bsn}
L.~Adams, E.~Chaubey, and S.~Weinzierl,
\newblock (2018), arXiv:1804.11144.

\bibitem{Adams:2018kez}
L.~Adams, E.~Chaubey, and S.~Weinzierl,
\newblock (2018), arXiv:1806.04981.

\bibitem{Sogaard:2014jla}
M.~Søgaard and Y.~Zhang,
\newblock Phys. Rev. {\bf D91}, 081701 (2015), arXiv:1412.5577.

\bibitem{Bonciani:2016qxi}
R.~Bonciani {\em et~al.},
\newblock JHEP {\bf 12}, 096 (2016), arXiv:1609.06685.

\bibitem{vonManteuffel:2017hms}
A.~von Manteuffel and L.~Tancredi,
\newblock JHEP {\bf 06}, 127 (2017), arXiv:1701.05905.

\bibitem{Primo:2017ipr}
A.~Primo and L.~Tancredi,
\newblock Nucl. Phys. {\bf B921}, 316 (2017), arXiv:1704.05465.

\bibitem{Ablinger:2017bjx}
J.~Ablinger {\em et~al.},
\newblock (2017), arXiv:1706.01299.

\bibitem{Bourjaily:2017bsb}
J.~L. Bourjaily, A.~J. McLeod, M.~Spradlin, M.~von Hippel, and M.~Wilhelm,
\newblock Phys. Rev. Lett. {\bf 120}, 121603 (2018), arXiv:1712.02785.

\bibitem{Hidding:2017jkk}
M.~Hidding and F.~Moriello,
\newblock (2017), arXiv:1712.04441.

\bibitem{Passarino:2017EPJC}
G.~{Passarino},
\newblock European Physical Journal C {\bf 77}, 77 (2017), arXiv:1610.06207.

\bibitem{Remiddi:2017har}
E.~Remiddi and L.~Tancredi,
\newblock Nucl. Phys. {\bf B925}, 212 (2017), arXiv:1709.03622.

\bibitem{Broedel:2017kkb}
J.~Broedel, C.~Duhr, F.~Dulat, and L.~Tancredi,
\newblock JHEP {\bf 05}, 093 (2018), arXiv:1712.07089.

\bibitem{Broedel:2017siw}
J.~Broedel, C.~Duhr, F.~Dulat, and L.~Tancredi,
\newblock Phys. Rev. {\bf D97}, 116009 (2018), arXiv:1712.07095.

\bibitem{Broedel:2018iwv}
J.~Broedel, C.~Duhr, F.~Dulat, B.~Penante, and L.~Tancredi,
\newblock (2018), arXiv:1803.10256.

\bibitem{Lee:2017qql}
R.~N. Lee, A.~V. Smirnov, and V.~A. Smirnov,
\newblock JHEP {\bf 03}, 008 (2018), arXiv:1709.07525.

\bibitem{Lee:2018ojn}
R.~N. Lee, A.~V. Smirnov, and V.~A. Smirnov,
\newblock (2018), arXiv:1805.00227.

\bibitem{Broedel:2014vla}
J.~Broedel, C.~R. Mafra, N.~Matthes, and O.~Schlotterer,
\newblock JHEP {\bf 07}, 112 (2015), arXiv:1412.5535.

\bibitem{Broedel:2015hia}
J.~Broedel, N.~Matthes, and O.~Schlotterer,
\newblock J. Phys. {\bf A49}, 155203 (2016), arXiv:1507.02254.

\bibitem{Broedel:2017jdo}
J.~Broedel, N.~Matthes, G.~Richter, and O.~Schlotterer,
\newblock J. Phys. {\bf A51}, 285401 (2018), arXiv:1704.03449.

\bibitem{DHoker:2015wxz}
E.~D'Hoker, M.~B. Green, {\"O}.~G{\"u}rdogan, and P.~Vanhove,
\newblock Commun. Num. Theor. Phys. {\bf 11}, 165 (2017), arXiv:1512.06779.

\bibitem{Hohenegger:2017kqy}
S.~Hohenegger and S.~Stieberger,
\newblock Nucl. Phys. {\bf B925}, 63 (2017), arXiv:1702.04963.

\bibitem{Broedel:2018izr}
J.~Broedel, O.~Schlotterer, and F.~Zerbini,
\newblock (2018), arXiv:1803.00527.

\bibitem{Vermaseren:1998uu}
J.~A.~M. Vermaseren,
\newblock Int. J. Mod. Phys. {\bf A14}, 2037 (1999), hep-ph/9806280.

\bibitem{Remiddi:1999ew}
E.~Remiddi and J.~A.~M. Vermaseren,
\newblock Int. J. Mod. Phys. {\bf A15}, 725 (2000), hep-ph/9905237.

\bibitem{Beilinson:1994}
A.~Beilinson and A.~Levin,
\newblock in {\it Motives}, ed. U. Jannsen, S. Kleiman, J.-P. Serre, Proc. of
  Symp. in Pure Mathematics {\bf 55}, Part 2, AMS, 1994, 123-190.

\bibitem{Levin:1997}
A.~Levin,
\newblock Comp. Math. {\bf 106}, 267 (1997).

\bibitem{Levin:2007}
A.~Levin and G.~Racinet,
\newblock (2007), arXiv:math/0703237.

\bibitem{Enriquez:2010}
B.~{Enriquez},
\newblock Selecta Math. {\bf 20}, 491 (2014), arXiv:1003.1012.

\bibitem{Brown:2011}
F.~Brown and A.~Levin,
\newblock (2011), arXiv:1110.6917.

\bibitem{Wildeshaus}
J.~Wildeshaus,
\newblock Lect. Notes Math. {\bf 1650}, Springer, (1997).

\bibitem{Kotikov:1990kg}
A.~V. Kotikov,
\newblock Phys. Lett. {\bf B254}, 158 (1991).

\bibitem{Kotikov:1991pm}
A.~V. Kotikov,
\newblock Phys. Lett. {\bf B267}, 123 (1991).

\bibitem{Remiddi:1997ny}
E.~Remiddi,
\newblock Nuovo Cim. {\bf A110}, 1435 (1997), hep-th/9711188.

\bibitem{Gehrmann:1999as}
T.~Gehrmann and E.~Remiddi,
\newblock Nucl. Phys. {\bf B580}, 485 (2000), hep-ph/9912329.

\bibitem{Argeri:2007up}
M.~Argeri and P.~Mastrolia,
\newblock Int. J. Mod. Phys. {\bf A22}, 4375 (2007), arXiv:0707.4037.

\bibitem{MullerStach:2012mp}
S.~M{\"u}ller-Stach, S.~Weinzierl, and R.~Zayadeh,
\newblock Commun.Math.Phys. {\bf 326}, 237 (2014), arXiv:1212.4389.

\bibitem{Henn:2013pwa}
J.~M. Henn,
\newblock Phys. Rev. Lett. {\bf 110}, 251601 (2013), arXiv:1304.1806.

\bibitem{Henn:2014qga}
J.~M. Henn,
\newblock J. Phys. {\bf A48}, 153001 (2015), arXiv:1412.2296.

\bibitem{Ablinger:2015tua}
J.~Ablinger {\em et~al.},
\newblock Comput. Phys. Commun. {\bf 202}, 33 (2016), arXiv:1509.08324.

\bibitem{Adams:2017tga}
L.~Adams, E.~Chaubey, and S.~Weinzierl,
\newblock Phys. Rev. Lett. {\bf 118}, 141602 (2017), arXiv:1702.04279.

\bibitem{Bosma:2017hrk}
J.~Bosma, K.~J. Larsen, and Y.~Zhang,
\newblock Phys. Rev. {\bf D97}, 105014 (2018), arXiv:1712.03760.

\bibitem{Kontsevich:2001}
M.~Kontsevich and D.~Zagier,
\newblock in: B. Engquist and W. Schmid, editors, Mathematics unlimited - 2001
  and beyond , 771 (2001).

\bibitem{Chen}
K.-T. Chen,
\newblock Bull. Amer. Math. Soc. {\bf 83}, 831 (1977).

\bibitem{Goncharov_no_note}
A.~B. Goncharov,
\newblock Math. Res. Lett. {\bf 5}, 497 (1998).

\bibitem{Goncharov:2001}
A.~B. Goncharov,
\newblock (2001), math.AG/0103059.

\bibitem{Borwein}
J.~M. Borwein, D.~M. Bradley, D.~J. Broadhurst, and P.~Lisonek,
\newblock Trans. Amer. Math. Soc. {\bf 353:3}, 907 (2001), math.CA/9910045.

\bibitem{Moch:2001zr}
S.~Moch, P.~Uwer, and S.~Weinzierl,
\newblock J. Math. Phys. {\bf 43}, 3363 (2002), hep-ph/0110083.

\bibitem{Vollinga:2004sn}
J.~Vollinga and S.~Weinzierl,
\newblock Comput. Phys. Commun. {\bf 167}, 177 (2005), hep-ph/0410259.

\bibitem{Brown:2014aa}
F.~{Brown},
\newblock (2014), arXiv:1407.5167.

\bibitem{Tarasov:1996br}
O.~V. Tarasov,
\newblock Phys. Rev. {\bf D54}, 6479 (1996), hep-th/9606018.

\bibitem{Tarasov:1997kx}
O.~V. Tarasov,
\newblock Nucl. Phys. {\bf B502}, 455 (1997), hep-ph/9703319.

\bibitem{Bogner:2010kv}
C.~Bogner and S.~Weinzierl,
\newblock Int. J. Mod. Phys. {\bf A25}, 2585 (2010), arXiv:1002.3458.

\bibitem{Bogner:2007mn}
C.~Bogner and S.~Weinzierl,
\newblock J. Math. Phys. {\bf 50}, 042302 (2009), arXiv:0711.4863.

\bibitem{Tkachov:1981wb}
F.~V. Tkachov,
\newblock Phys. Lett. {\bf B100}, 65 (1981).

\bibitem{Chetyrkin:1981qh}
K.~G. Chetyrkin and F.~V. Tkachov,
\newblock Nucl. Phys. {\bf B192}, 159 (1981).

\bibitem{vanHoeij:1997}
M.~van Hoeij,
\newblock J. Symbolic Computation {\bf 24}, 537 (1997).

\bibitem{Baikov:1996iu}
P.~A. Baikov,
\newblock Nucl. Instrum. Meth. {\bf A389}, 347 (1997), arXiv:hep-ph/9611449.

\bibitem{Lee:2009dh}
R.~N. Lee,
\newblock Nucl. Phys. {\bf B830}, 474 (2010), arXiv:0911.0252.

\bibitem{Kosower:2011ty}
D.~A. Kosower and K.~J. Larsen,
\newblock Phys. Rev. {\bf D85}, 045017 (2012), arXiv:1108.1180.

\bibitem{CaronHuot:2012ab}
S.~Caron-Huot and K.~J. Larsen,
\newblock JHEP {\bf 1210}, 026 (2012), arXiv:1205.0801.

\bibitem{Frellesvig:2017aai}
H.~Frellesvig and C.~G. Papadopoulos,
\newblock JHEP {\bf 04}, 083 (2017), arXiv:1701.07356.

\bibitem{Bosma:2017ens}
J.~Bosma, M.~Sogaard, and Y.~Zhang,
\newblock JHEP {\bf 08}, 051 (2017), arXiv:1704.04255.

\bibitem{Harley:2017qut}
M.~Harley, F.~Moriello, and R.~M. Schabinger,
\newblock JHEP {\bf 06}, 049 (2017), arXiv:1705.03478.

\bibitem{Meyer:2016slj}
C.~Meyer,
\newblock JHEP {\bf 04}, 006 (2017), arXiv:1611.01087.

\bibitem{Meyer:2017joq}
C.~Meyer,
\newblock Comput. Phys. Commun. {\bf 222}, 295 (2018), arXiv:1705.06252.

\bibitem{Stein}
W.~A. Stein,
\newblock {\em Modular Forms, a Computational Approach} (American Mathematical
  Society, 2007).

\bibitem{Bognerandmore}
C.~Bogner, A.~Schweitzer, and S.~Weinzierl,
\newblock these proceedings.

\end{thebibliography}
\end{document}